\documentclass[a4paper,10pt,english]{article}

\usepackage{helvet}         
\usepackage{courier}        
\usepackage{type1cm}        
\usepackage{makeidx}         
\usepackage{graphicx}        
\usepackage{multicol}        
\usepackage[bottom]{footmisc}
\usepackage{babel}
\RequirePackage{doi}

\usepackage{amssymb}
\usepackage{cite}
\usepackage{mathtools}
\usepackage{empheq}
\usepackage{url}
\usepackage{comment}
\usepackage[mathscr]{euscript}
\usepackage[all,color]{xy}
\usepackage{relsize}
\usepackage{url}
\usepackage{cancel}
\usepackage{caption}
\usepackage[T1]{fontenc}
\usepackage{color}
\usepackage{mathtools}

\RequirePackage{hyperref}

\newcommand{\bbm}{\left(\begin{matrix}}
    \newcommand{\ebm}{\end{matrix}\right)}
\newcommand{\beq}{\begin{eqnarray}}
\newcommand{\eeq}{\end{eqnarray}}

\makeatother

 \def\one{\mbox{1 \kern-.59em {\rm l}}}

\begin{document}

\title{Intertwining noncommutativity with gravity and particle physics}

\author{G Manolakos$^1$, P Manousselis$^2$, D Roumelioti$^2$, S Stefas$^2$, G Zoupanos$^{2,3,4,5}$}\date{}
\maketitle

\begin{center}
\itshape$^1$ Institute of Theoretical Physics, Faculty of Physics, University of Warsaw, ul. Pasteura 5, Warsaw 02-093, Poland\\
\itshape$^2$Physics Department, National Technical University, Athens, Greece\\
\itshape$^3$ Theory Department, CERN\\
\itshape$^4$ Max-Planck Institut f\"ur Physik, M\"unchen, Germany\\
\itshape$^5$ Institut f\"ur Theoretische Physik der Universit\"at Heidelberg, Germany
\end{center}

\begin{center}
\emph{E-mails: \href{mailto:giorgismanolakos@gmail.com}{giorgismanolakos@gmail.com}, \href{mailto:pman@central.ntua.gr}{pman@central.ntua.gr},\\ \href{mailto:danai\_roumelioti@mail.ntua.gr}{danai\_roumelioti@mail.ntua.gr}, \href{mailto:dstefas@mail.ntua.gr}{dstefas@mail.ntua.gr}, \href{mailto:George.Zoupanos@cern.ch}{George.Zoupanos@cern.ch}}
\end{center}
\begin{abstract}
    Here we present an overview on the various works, in which many collaborators have contributed, regarding the interesting dipole of noncommutativity and physics. In brief, we present the features that noncommutativity triggers both in the fields of gravity and particle physics, from a matrix-realized perspective, with the notion of noncommutative gauge theories to play the most central role in the whole picture. Also, under the framework of noncommutativity, we examine the possibility of unifying the two fields (gravity-particle physics) in a single configuration.     
\end{abstract}

\section{Introduction}


An ultimate anticipation of many theoretical physicists is the existence of a unification picture in which all fundamental interactions are involved. To this end, a huge amount of serious research activity has been carried out, including works that elaborate the very interesting notion of extra dimensions. Superstring theories \cite{green-scwarz-witten} consist a solid framework, with the heterotic string theory \cite{gross-harvey} (defined in ten dimensions) being the most promising, in the sense that it potentially admits experimental compatibility, due to the fact that the Standard Model (SM) gauge group can be accommodated into the gauge groups of the Grand Unified Theories (GUTs) that emerge after the dimensional reduction of the initial $E_8\times E_8$. Besides the superstring theories, a few years before their formulation, an alternative approach of generalized dimensional reduction, as compared to the simple one of higher-dimensional gauge theories was formulated. This insightful and significant project which shared common goals with one of the superstring theories, was initially explored by Forgacs-Manton and then by Scherk-Schwartz studying the Coset Space Dimensional Reduction (CSDR) \cite{forgacs-manton,kapetanakis-zoupanos, kubyshin-mourao, Manousselis:2000aj} and the group manifold reduction \cite{scherk-schwarz}, respectively.

Besides the above, a very interesting framework which could be a competitive candidate in accommodating a gravitational theory but also particle physics models at high energy scale (Planck scale) is that of noncommutative geometry \cite{connes, madorej, Madore:1991bw, buric-grammatikopoulos-madore-zoupanos, filk, Varilly:1998gq, Chaichian:1998kp, Minwalla:1999px, grosse-wulkenhaar, grosse-steinacker, Grosse:2004wm, connes-lott, Chamseddine:1996zu, Chamseddine:2007ia, martin-bondia, dubois-madore-kerner, DuboisViolette:1988ps, DuboisViolette:1988vq, madorejz, Madore:1992ej, connes-douglas-schwarz, seiberg-witten, ishibasi-kawai, jurco, Jurco:2001my, Jurco:2001rq, Barnich:2002pb, chaichian, camlet, Aschieri:2002mc, Behr:2002wx, aschieri-madore-manousselis-zoupanos, Aschieri:2004vh, Aschieri:2005wm, Aschieri:2006uw, Aschieri:2007fb, Steinacker:2007ay, Chatzistavrakidis:2009ix, Chatzistavrakidis:2010xi, Chatzistavrakidis:2010tq, Gavriil:2015lka, Manolakos:2017nts, Manolakos:2016wpx}, in which the commutativity of the coordinates is not an inherent property of spacetime. An interesting virtue of the above framework is the regularization of quantum field theories, or, even better, the construction of finite ones. Of course, such an undertaking is rather complicated and, furthermore, it has presented unwelcome issues regarding its ultraviolet behavior \cite{filk, Varilly:1998gq, Chaichian:1998kp, Minwalla:1999px}(see also \cite{grosse-wulkenhaar, grosse-steinacker, Grosse:2004wm, Sitarz}). Despite that, noncommutative geometry is considered as a solid framework regarding the accommodation of particle physics models, formulated as in the familiar way, that is as gauge theories on noncommutative spaces \cite{connes-lott, Chamseddine:1996zu, Chamseddine:2007ia}(see also \cite{martin-bondia, dubois-madore-kerner, DuboisViolette:1988ps, DuboisViolette:1988vq, madorejz, Madore:1992ej}).

It is remarkable that superstring theories and noncommutative geometry share a common ground, since, in M-theory and open string theory, the effective physics on D-branes can be expressed as a noncommutative gauge theory, in the presence of a nowhere-vanishing background antisymmetric field \cite{connes-douglas-schwarz, seiberg-witten}. In addition, the type IIB superstring theory (and those related to it through dualities), in its non-perturbative formulation as a matrix model \cite{ ishibasi-kawai}, consists a noncommutative theory. A major contribution in the framework of noncommutativity was that of Seiberg and Witten \cite{seiberg-witten}, who devised a mapping mechanism between commutative and noncommutative gauge theories, correlating the noncommutative degrees of freedom of a gauge theory to their commutative counterparts. Based on the above mapping, important developments have been done \cite{jurco, Jurco:2001my, Jurco:2001rq, Barnich:2002pb, chaichian}, for instance the construction of a noncommutative version of the SM \cite{camlet, Aschieri:2002mc, Behr:2002wx}. Unfortunately, the main problem of this approach was not solved, that is the free parameters of the model could not get further reduced by extensions of this type (contrary to supersymmetric theories \cite{heinemeyer}).


Delving a little deeper in the notion of noncommutative geometry, since coordinates are not commutative quantities, it can be linked to a potential quantum structure, which can be supposed that it occurs at very small distances (Planck length), since the behavior of the spacetime fabric at these scales is effectively unknown. According to the above line of thoughts, it is a rather natural step to examine the noncommutative version of General theory of Relativity (GR), with aspirations that the latter would provide new insights particularly in regions that a spacetime singularity is encountered in the conventional GR framework. This noncommutative gravitational theory would consist a generalization of GR, ideal for examining higher curvature scales (than a critical one), in which, localization of a point would be impossible to occur. Therefore, in case of high-scales phenomena, the conventional notion of coordinates breaks down and should be substituted by elements of a noncommutative algebra. On the contrary, at less extreme energy scales (e.g. LHC) the rest of the interactions are successfully formulated using gauge theories, while at higher scales (but not of Planck level) a very attractive unification picture of these three interactions is provided by the GUTs. The gravitational interaction does not join this picture, since, in principle, it is formulated geometrically according to GR. Nevertheless, besides the geometric one, there exists an alternative, gauge-theoretic  approach to gravity  \cite{Utiyama:1956sy, Kibble:1961ba, Stelle:1979aj, MacDowell:1977jt, Ivanov:1980tw, Ivanov:1981wm, Kibble:1985sn, Kaku:1977pa, Fradkin:1985am, vanproeyen, cham-thesis, Chamseddine:1976bf, Witten:1988hc}, which started with Utiyama's pioneering study \cite{Utiyama:1956sy} and was subsequently evolved as a gauge theory of the de Sitter SO(1,4) group, spontaneously broken by a scalar field to the Lorentz SO(1,3) group \cite{Stelle:1979aj}. Apart from GR, Weyl gravity has also been formulated as a gauge theory of the conformal group in four dimensions \cite{Kaku:1977pa, Fradkin:1985am}. In this case, part of the gauge fields spectrum is identified as the vielbein and the spin connection, which guarantees the interplay between the gauge-theoretic and geometric approaches through the (equivalent to the regular -second order-) first order formulation of GR. Now, taking into consideration the aforementioned gauge-theoretic formulation of gravity and integrating it to the noncommutative framework in which gauge theories are well-formulated has led to the construction of models of noncommutative gravity \cite{Chamseddine:2000si, Chamseddine:2003we, Aschieri1, Aschieri2, Aschieri_2005, Ciric:2016isg, Cacciatori:2002gq, Cacciatori:2002ib, Aschieri3, Banados:2001xw}. In these works, the Moyal-Weyl type of noncommutativity is used, the star-product approach is followed (in which the noncommutative quantities are still ordinary functions but with an upgraded product) and, last, the Seiberg-Witten map is used \cite{seiberg-witten}. Furthermore, there exists an alternative approach to the construction of noncommutative gravitational models, which makes use of the matrix-realization of the noncommutative quantities \cite{ishibasi-kawai, Aoki:1998vn, Hanada:2005vr, Furuta:2006kk, Yang:2006dk, Steinacker:2010rh, Kim:2011cr, Nishimura:2012xs,Nair:2001kr, Abe:2002in, Valtancoli:2003ve, Nair:2006qg,Banks:1996vh} (see also \cite{buric-grammatikopoulos-madore-zoupanos, Buric:2007zx, Buric:2007hb}).

Here, due to length restrictions of our article, instead of outlining our approach in the introduction, we have chosen to develop it step-by-step in the next chapters, accompanied by more detailed references on the subject.

\section{Gauge theories on Noncommutative space}

As explained in the introduction, in order to accommodate the gravitational interaction into the framework of noncommutative geometry, following \cite{Madore:2000en}, it is meaningful to recall how gauge theories are rephrased in it.  

Let a scalar field, $\phi(X)$, where $X$ consists the coordinate system of the noncommutative space. The scalar field transforms non-trivially under a local (infinitesimal) gauge transformation as:
\begin{equation}
    \delta \phi (X) = \varepsilon (X) \phi (X),
\end{equation}
where $\varepsilon(X)$ is the coordinate-dependent parameter of the transformation. Contrary to the scalar field, the coordinates transform trivially under gauge transformation (as expected), therefore the product $X_\mu \phi(X)$ transforms as:
\begin{equation}
    \delta(X_{\mu} \phi (X))= X_{\mu} \varepsilon (X) \phi(X)\,.
\end{equation}
By observation and taking into consideration that the underlying framework is the noncommutative one, it is understood that the above transformation is not a covariant one. In order to covariantize it, drawing lessons from the conventional gauge theories, the covariant coordinate is utilized (in analogy to the covariant derivative), the definition of which is given through its transformation:
\begin{equation}
    \delta \mathcal{X}_{\mu}=\left[\varepsilon(X), \mathcal{X}_{\mu}\right]\,,
\end{equation}
which is obtained by the requirement of the quantity $\delta(X_{\mu} \phi (X))$ to transform covariantly, that is
\begin{equation}
\delta\left(\mathcal{X}_{\mu} \phi(X)\right) \equiv \varepsilon(X) \mathcal{X}_{\mu} \phi(X)\,.
\end{equation}
From the above configuration, in order to relate the noncommutative coordinate, $X_\mu$, to the noncommutative covariant coordinate, $\mathcal{X}_\mu$, a field $\mathcal{A}_\mu$ with transformation
\begin{equation}
\label{deltaAlpha}
\delta \mathcal{A}_{\mu}(X)=-\left[X_{\mu}, \varepsilon(X)\right]+\left[\varepsilon(X), \mathcal{A}_{\mu}(X)\right]
\end{equation}
has to be introduced, specifically according to the relation $\mathcal{X}_\mu=X_\mu+\mathcal{A}_\mu$. It is clear from the latter that the introduced field $\mathcal{A}_\mu$ admits the interpretation of the noncommutative gauge connection. In turn, due to the above interpretation of the $\mathcal{A}_\mu$, a noncommutative version of a field strength tensor is corresponded in accordance to the ordinary gauge theories (including an extra term to its definition for reasons of covariance), which is dependent on the kind of noncommutative space on which the gauge theory is constructed.    

A rather important feature of the noncommutative gauge theories, in discordance to the conventional gauge theories, is that the anticommutators of the various operators related to the gauge algebra become relevant. This feature is designated when one considers the commutator of two elements which belong to the gauge algebra,  $\varepsilon(X)=\varepsilon^{a}(X) T_{a}$ and $\phi(X)=\phi^{a}(X) T_{a}$:
\begin{equation}
\label{anticom}
[\varepsilon, \phi]=\frac{1}{2}\left\{\varepsilon^{a}, \phi^{b}\right\}\left[T_{a}, T_{b}\right]+\frac{1}{2}\left[\varepsilon^{a}, \phi^{b}\right]\left\{T_{a}, T_{b}\right\},
\end{equation}
where $T_a$ denote the algebra generators. The quantities $\varepsilon^a$ and $\phi^b$ are functions of the spacetime coordinates, which means that, in the commutative case, their corresponding commutator vanishes, and subsequently so does the product of the last term in the above relation. Nevertheless, in the noncommutative case, by definition, the coordinates do not commute with each other and thus so do functions that depend on them. Therefore, the aforementioned last term of the above relation becomes non-vanishing giving the anticommutators an essential role in the construction of gauge theories in the noncommutative setting, contrary to the conventional ones. This discrepancy gives rise to terms that originate from the anticommutators, which, in principle, are not members of the gauge algebra. Consequently the closure property of the initial algebra holds no more. One way out is to extend the algebra perpetually including all operators that pop from the anticommutators, with this extension leading to the universal enveloping algebra, which, although useful in other contexts (e.g., in \cite{jurco, Aschieri2, dimitrijevic}), in ours it is not the appropriate way to proceed. Another way out is to restrict the number of the newly-added operators to a finite (minimum) number by choosing a specific representation, which is the one preferred. 

Now, let us briefly emphasize on the category of the covariant noncommutative spaces on which we focus \cite{Snyder:1946qz, Yang:1947ud, Madore:1991bw, Grosse:1993uq,Heckman:2014xha, Buric:2015wta, Buric:2017yes}, which have the property that Lorentz covariance is preserved \cite{Kimura:2002nq, Steinacker:2016vgf, Sperling:2017dts, yang:2009}. Moreover, another property of the noncommutative spaces is the preservation of the isometries of the corresponding commutative space. The noncommutative spaces forming this special subclass are called fuzzy and the most typical example is the fuzzy 2-sphere \cite{Madore:1991bw} (see also \cite{dolan:2002, oconnor:2006}) which shares the same isometry group, $SO(3)$, with its commutative analogue, i.e. the ordinary sphere. The fuzzy sphere admits a construction through finite-dimensional matrices, the size of which is interpreted as the number of the quanta of the space. The coordinates of the fuzzy sphere with $N-1$ level of fuzziness are $N\times N$ matrices which are multiples of the $SU(2)$ generators in the $N-$dimensional representations. The aforementioned construction is deployed through the truncation of the angular momentum by the introduction of a cut-off parameter $N-1$ and therefore leaving $N^2$ independent functions. Thus, the above functions may be represented by $N\times N$ matrices leading to a noncommutative algebra of the sphere.

Attempting to generalize the above argument to the four-dimensional case, i.e. the one that interests us, it is seen that such a mapping of functions to matrices cannot be achieved, as the number of independent functions does not coincide with the square of some integer number. Therefore, the construction of the fuzzy 4-sphere and field theories on it is a more subtle task. For alternative constructions see \cite{Kimura:2002nq} and referenced studies in it (see also \cite{medina:2012, medina:2003}).   

Overall, it has already become clear that, according to our (and others') perspective and approach, gauge theories play an important role in the construction of both particle physics models as well as gravitational ones in the framework of noncommutativity. In the ensuing, we will present constructions that are related to both, starting with the particle physics side followed by the gravitational one. 

\section{Fuzzy particle physics model}
In this section we describe a fuzzy particle physics model in which fuzziness is introduced through the notion of extra dimensions. The latter are not supposed to be in an arbitrary form but they rather consist specific (extra-dimensional) manifolds, particularly noncommutative analogues (fuzzy) of the well-defined and well-studied coset spaces. Before we move on with the discussion of the specific model, it is rather considerate first to write down some information of the backbone of the whole construction, that is the dimensional reduction of a higher-dimensional gauge theory with fuzzy extra dimensions, in the most general case, in which the extra dimensions consist an arbitrary fuzzy coset space.    

\subsection{Dimensional Reduction of a Higher-Dimensional Theory with Fuzzy Extra Dimensions}


Suppose a higher-dimensional theory defined on the spacetime $M^4 \times (S/R)_F$ of $D = d+4$ dimensions, where $(S/R)_F$ is a fuzzy coset space. Let this theory be gauge invariant under the transformations of the group $G=U(P)$, with generators $\mathcal{T}^I$ and be described by the following action of Yang-Mills type:
\begin{equation}
\label{ymaction}
    \mathcal{S}_{\text{YM}}=\frac{1}{4g^2}\int d^4 x\, k \operatorname{Tr} \operatorname{tr}_G F_{MN}F^{MN}\,,
\end{equation}
where $\operatorname{tr}_G$ is the trace related to the generators of the algebra of the gauge group $G$ and $k \operatorname{Tr}$. Also, although the theory is higher-dimensional, the integration is written in such a way as it is performed on the four-dimensional component. The correct dimensions regarding the integration are restored by the presence of the $\operatorname{Tr}$ which is effectively the integration operator referring to the coordinates of the extra dimensions, since they now are represented by matrices. Moreover, $k$ is a parameter related to the volume of the fuzzy coset space (see \cite{manolakos2019} for the case of the fuzzy sphere) and $F_{MN}$ is the field strength tensor of the gauge theory with $M, N = 0,\ldots, D-1$. That means that the tensor $F_{MN}$ consists of components which lie only on the four-dimensional space, components that lie only on the extra-dimensional one, as well as mixed components of the two spaces, that is $F_{MN} = (F_{\mu \nu}, F_{\mu b}, F_{a\nu}, F_{ab})$, where $\mu, \nu = 0,\ldots 3$ and $a, b = 4, \ldots D-1$. Specifically, the mixed and fuzzy parts of the tensor are:
\begin{equation*}
\begin{gathered}
     F_{\mu a} = \partial_\mu \phi_a + [A_\mu , \phi_a] = D_\mu \phi_a\\
    F_{ab}=[X_a, A_b]-[X_b, A_a]+[A_a , A_b ] - C^c{}_{ba} A_{ac},     
\end{gathered}
\end{equation*}
where $\phi_a$ stands for the covariant coordinate, that is the noncommutative analogue of the covariant derivative (see section \ref{gravity} for a more detailed discussion). The initial action \eqref{ymaction}, after taking into consideration the above two expressions of the components of the field strength tensor, takes the following form:

\begin{equation}
\label{ymaction2}
    \mathcal{S}_{\text{YM}}=\int d^4 x \operatorname{Tr}\operatorname{tr}_G\left(\frac{k}{4g^2}F^2_{\mu\nu}+\frac{k}{2g^2}(D_\mu \phi_a)^2-V(\phi)\right),
\end{equation}
where $V(\phi)$ is a function that involves the terms that originate from the $F^2_{ab}$ one:
\begin{equation}
\begin{aligned}
    V(\phi)&= -\frac{k}{4g^2} \operatorname{Tr}\operatorname{tr}_G \sum_{ab} F_{ab}F_{ab}\\
           &=-\frac{k}{4g^2} \operatorname{Tr}\operatorname{tr}_G\left([\phi_a, \phi_b] [\phi^a ,\phi^b]-4C_{abc} \phi^a \phi^b \phi^c + 2R^{-2} \phi^2\right).
\end{aligned}
\end{equation}
Due to its expression, the above is identified as the potential of the theory. The form in which the action arrives after the above substitutions of the field strength tensor according to the splitting we considered in its components, \eqref{ymaction2}, manifestly lets us naturally interpret it as a four-dimensional action. The above interpretation is also possible at the level of the gauge transformation; let $\lambda(x^\mu , X^a $) be an infinitesimal parameter of a local transformation of the initial gauge group, $G =U(P)$. It admits the interpretation of a local transformation of another group on a gauge theory exclusively on $M^4$, if treated in the following way:
\begin{equation}
\label{lambda}
    \lambda(x_\mu , X_a ) = \lambda^I (x^\mu , X^a )\mathcal{T}^I = \lambda^{h,I}(x^\mu )\mathcal{T}^h\mathcal{T}^I\,,
\end{equation}
where, as mentioned already, $\mathcal{T}^I$ are the (Hermitian) generators of the gauge group $U(P)$ of the initial higher-dimensional theory and $\lambda^I (x_\mu , X^a)$ are $N\times N$ anti-Hermitian matrices, i.e. functions of the matrix-realized coordinates. Since we are now dealing with $N\times N$ antisymmetric matrices, $\lambda^I (x_\mu , X^a)$, they can be thought as an element of the $U(N)$ group and, as such, they can be written down as decompositions on the corresponding generators, $\mathcal{T}^h$, as $\lambda^I (x_\mu , X^a) = \lambda^{h,I}(x^\mu )\mathcal{T}^h$, where $\lambda^{h,I}(x^\mu)$ being merely functions of the four-dimensional coordinates identified as the Kaluza-Klein modes of $\lambda^I (x_\mu , X^a )$ and, they can be further understood as a field that takes values in the Lie algebra of the $U(N)\times U(P)$ group, which is equivalently the algebra of the $U(NP)$ group. Also, under same treatment, the gauge connection of the initial gauge theory can be viewed from a four-dimensional perspective as a four-dimensional one of the $U(NP)$ group. Last, the same applies on the scalars as well.

The above reduction, although straightforward and rather simple, gives rise to a very important and anti-intuitional feature (which can also be encountered in more complicated ones), that is the gauge symmetry of the four-dimensional theory is larger than that of the initial, higher-dimensional one. In other words, one may start with a higher-dimensional Abelian gauge theory and end up with a non-Abelian in four dimensions. Again from the above reduction, it is also understood that the scalars belong to the adjoint representation of the four-dimensional gauge group, as relics of the initial gauge fields which are also in the corresponding adjoint, fact that leads to the understanding that they cannot trigger the electroweak symmetry breaking. This is a crucial reason why the above simple model cannot be evolved into a more complicated and promising one and consequently, another more elaborate one has to be sought. 

Following the above consideration, an improved four-dimensional model may be obtained by employing a fuzzy version \cite{Aschieri:2003vy, Aschieri:2004vh, Aschieri:2005wm} of the CSDR \cite{Kapetanakis_1993}. Due to the fuzziness, the fuzzy CSDR will enjoy the above welcome feature that was noted above, that of the enlargement of the gauge symmetry as the number of dimensions drops to four, contrary to the conventional CSDR in which this feature is absent. An important observation is that in the fuzzy case choosing an Abelian gauge theory in high dimensions one is led to non-Abelian gauge symmetry in four dimensions. Another virtue the fuzzy CSDR inherits from the fuzziness is that both the higher-dimensional and the resulting four-dimensional theories are renormalizable. 

The matter of renormalizability was initially argued in \cite{Aschieri:2003vy, Aschieri:2004vh} but an even more convincing argument came after the whole problem was examined from another perspective. In a few words, instead of beginning with a higher-dimensional theory and performing a dimensional reduction to approach a four-dimensional theory, the starting point was overturned, in the sense that one may begin with a renormalizable four-dimensional gauge theory of $SU(N)$ with scalars populating a multiplet in such a way that fuzzy extra dimensions forming a fuzzy sphere can be developed dynamically \cite{Aschieri:2006uw}. Such a theory can be interpreted as a six-dimensional gauge theory since it develops non-trivial vacua, with the geometrical and gauge symmetry determined by the parameters of the initial Lagrangian. Also, the tower of (massive) Kaluza-Klein modes that emerges is finite, in consistency with the view of a compactified gauge theory in higher dimensions. Briefly, the virtues of the above model are, first, that, avoiding supersymmetry or fine tuning, extra dimensions emerge dynamically due to the fact that fuzzy spaces consist solutions of matrix models. Second, the four-dimensional gauge group is $SU(n_1) \times SU(n_2) \times U(1)$ or $SU(n)$, while gauge groups consisting of more than two simple groups are not observed in this kind of models. Third, the induction of a magnetic flux occurs in a rather natural manner in the case of vacua having non-simple gauge symmetry.    

Subsequently, these nice features of the above mechanism suggest that it is worth attempting to accommodate particle physics models with phenomenological orientation. To this end, chiral fermions are included and relevant studies have shown that, without imposing any additional constraints, best case scenario is that of obtaining four-dimensional theories populating bi-fundamental representations (mirror fermions) \cite{Steinacker:2007ay}, \cite{Chatzistavrakidis:2009ix}. Although having mirror fermions is not a killing result for phenomenological compatibility \cite{ MAALAMPI199053}, chiral fermions are significantly preferred. Such an outcome occurs after an additional mechanism is applied in the above, specifically that of a $\mathbb{Z}_3$ orbifold projection of an $\mathcal{N} = 4$ supersymmetric $SU(3N)$ gauge theory which eventually leads to an $\mathcal{N}=1$ $SU(N)^3$ gauge theory \cite{Chatzistavrakidis:2010xi}. The case of $N=3$ (trinification group) is of particular interest (see \cite{Heinemeyer_2010, Ma_2004, Heinemeyer_2010_2}). 

Let us see in some detail the above concepts through a particle physics model in which fuzzy extra dimensions are dynamically generated to form fuzzy spheres and chiral fermions are involved due to the orbifold projection mechanism \cite{Chatzistavrakidis:2010xi}. 

As mentioned earlier, let us consider an $\mathcal{N}=4$ Supersymmetric Yang-Mills (SYM) theory in four dimensions with gauge symmetry parametrized by $SU(3N)$. The particle spectrum consists of the gauge field, $A_\mu$, three complex scalars, $\phi^i$, and four Majorana fermions $\psi^p$ all in the adjoint representation of the gauge group. The orbifold projection is parametrized by the action of a $\mathbb{Z}_3$ discrete group and is realized through its embedding into the R-symmetry, that is $SU(4)_R$. The choice of the embedding is not unique and this choice determines the amount of the resulting sypersymmetry \cite{Kachru_1998}. Here, the discrete group is considered to get embedded into the $SU(3)$ subgroup of the R-symmetry, a choice which leads to $\mathcal{N}=1$ remnant supersymmetry, breaks the gauge symmetry to $SU(N)\times SU(N)\times SU(N)$ and the only fields that make it to the resulting SYM theory are the ones that are $\mathbb{Z}_3$ invariant. It should be noted that the fermions reside in chiral representation of the resulting group and come in three identical copies, that is three chiral families. 

At the level of the initial $\mathcal{N}$ SYM theory the F-part of the scalar potential is:
\begin{equation}
    V_F(\phi)=\frac{1}{4}\operatorname{Tr}\left([\phi^i,\phi^j]^{\dagger}[\phi^i,\phi^j]\right) 
\end{equation}
and its form remains the same after the orbifold projection filters the spectrum. The D-part of the potential is $V_D=\frac{1}{2}D^2=\frac{1}{2}D^ID_I$, where $D^I=\phi_i^{\dagger}T^I\phi^i$ with $T^I$ the generators of the gauge group in chiral representations. Minimization of the total potential gives $[\phi^i,\phi^j]=0$. However, the introduction of soft supersymmetric breaking terms with scalar part which respects the orbifold symmetry:
\begin{equation}
    V_{SSB}=\frac{1}{2}\sum_i m_i^2 {\phi^i}^{\dagger}\phi^i+\frac{1}{2}\sum_{i,j,k}h_{ijk}\phi^i\phi^j\phi^k+\operatorname{h.c.}\,,
\end{equation}
contributes in such a way to the minimization of the potential that leads to different kind of vacua. The potential of the combination of all three contributions is given in the following form:
\begin{equation}
    V=\frac{1}{4}({F^{ij})}^{\dagger}F^{ij}+V_D\,,
\end{equation}
for suitable parameters, where:
\begin{equation}
    F^{ij}=[\phi^i,\phi^j]-i\varepsilon^{ijk}{(\phi^k)}^{\dagger}.
\end{equation}
The first term of the above potential is positive definite, therefore its global minimum is obtained if the following equations hold:
\begin{equation}
    [\phi^i,\phi^j]=i\varepsilon_{ijk}{(\phi^k)}^{\dagger},\quad
    [{(\phi^i)}^{\dagger},{(\phi^j)}^{\dagger}]=i\varepsilon_{ijk}\phi^k,\quad
    \phi^i{(\phi^i)}^{\dagger}=R^2\,,
\end{equation}
where ${(\phi^i)}^{\dagger}$ is the hermitian conjugate of $\phi^i$ and it holds that $[R^2,\phi^i]=0$. The above relations point towards the fuzzy sphere defining relation, which becomes even more transparent by considering the (untwisted) complex scalar fields, $\Tilde{\phi^i}$, which are defined as $ \phi^i=\Omega\title{\phi^i}\,,$
for $\Omega\neq 1$, $\Omega^3=1$, $[\Omega,\phi^i]=0$, $\Omega^{\dagger}=\Omega^{-1}$ and ${{(\tilde{\phi}^i)}}^{\dagger}=\Tilde{\phi^i}$, i.e. ${(\phi^i)}^{\dagger}=\Omega\phi^i$.

To conclude, fuzzy extra dimensions equipped with orbifold projection consist a valuable asset when it comes to particle physics models attaching importance to them in the aspects of chirality, renormalizability and phenomenological viability. 

\section{Noncommutative gravity}
\label{gravity}
As mentioned in the introduction, gauge theories in the noncommutative framework are employed for constructing both particle physics and gravitational models. Having reviewed the first part, from now on we will examine the second one in two cases, that of three and four dimensions. 
\subsection{Fuzzy gravity in three dimensions on the \texorpdfstring{$\mathbb{R}_{\lambda}^3$}{R lambda 3} space}
First we review a matrix-realized description of a gauge-theoretic construction of noncommutative gravity in three dimensions. For that to occur, a three-dimensional noncommutative space is required as a background in order to accommodate the whole construction. Therefore, we start our discussion with the description of that three-dimensional space.

\subsection*{\textit{The $\mathbb{R}_{\lambda}^3$ space}}
The path of the description of the noncommutative space that is used in this case passes through the fuzzy sphere, which is a very fundamental and well-defined fuzzy space that is also equipped with the property of covariance \cite{Madore:1991bw, Hoppe}. A glimpse at its properties has  already been given in the setup of the particle models of the previous section as it had a very important role in the corresponding construction as well. Moreover, the definition of the fuzzy sphere is given by the commutation relation the coordinates satisfy and the radius constraint, that is:
\begin{equation}
\label{coordcomm}
    [X_i,X_j]=i\lambda \epsilon_{ijk}X_k~, \quad \sum_{i=1}^3 X_i X_i=\lambda^2 j(j+1):=r^2\,.
\end{equation}
The first relation is actually a rescaled version of the algebra the angular momentum operators satisfy due to the identification $X_i=\lambda J_i$, where $\lambda$ fixes the dimensions and $J_i$ operators are in a unitary irreducible (high) representation of $SU(2)$. The second relation is a rescaled version of the Casimir operator which is manifestly interpreted as the radius constraint of the fuzzy sphere. The above definition of the fuzzy sphere leads smoothly to the definition of the space that is used as background for the construction of the gauge theory, namely $\mathbb{R}_{\lambda}^3$ \cite{Hammou:2001cc, Vitale:2012dz, Vitale:2014hca, Wallet_2016}, if the radius constraint is switched off by considering unitary reducible representations for the generators. As known, a reducible representation may be written in a block diagonal form of irreducible representations, or to rephrase, a block diagonal form of fuzzy spheres. This property of $\mathbb{R}_{\lambda}^3$ gives rise to a very illustrative view, that is the visualization of the above space as a foliation of the three-dimensional Euclidean space by multiple fuzzy spheres of all possible radii, according to the number of the fuzziness level \cite{DeBellis}.  

\subsection*{\textit{Gauge theory of three-dimensional gravity on $\mathbb{R}_{\lambda}^3$}}
Now, having determined the background three-dimensional space, a gravitational model can be realized by considering the gauge theory of a suitable group on it \cite{Chatzistavrakidis:2018vfi}.  

The above methodology is a loan from the formulation of three-dimensional GR as a Chern-Simons gauge theory on the three-dimensional Minkowski spacetime background. Specifically, the gauge group that successfully fits in the above configuration is the isometry group of the background space, that is the three-dimensional Poincar\'e group, $ISO(1,2)$ \cite{Witten:1988hc}. As noted in the introduction, the above translation of GR to the gauge-theoretic setting is achieved by considering the first order formulation of gravity, that is the formulation in which dynamics is described by the vielbein and the spin connection instead of the metric. A very important consideration is that the latter two dynamical quantities are integrated in the gauge-theoretic setting as gauge fields related to the translational and Lorentz part of the Poincar\'e group, respectively. 

Back in the noncommutative framework, the first thing that has to be determined, as understood by the above information, is the gauge group. Along the lines of the commutative case, the one that is employed is again the isometry group of the background space $\mathbb{R}_{\lambda}^3$, that is $SO(4)$, which leads to the construction of a non-Abelian noncommutative gauge theory. As pointed out in section 2, such theories complicate things as anti-commutators become important and the way out is to fix the representation and extend the group by those operators that the anticommutators give. The resulting gauge theory is that of the $U(2)\times U(2)$ and the corresponding generators are represented by $4\times 4$ matrices (for details but also for the manipulation of the Lorentzian case see the original publication \cite{Chatzistavrakidis:2018vfi}).    

The set of the eight generators of the resulting gauge group is: $\{P_a,~ M_a,~ \mathbb{I},~ \gamma_{5}\}$, where $a = 1,2,3$ and their specific form is given after employing the Pauli matrices of which the commutation and anticommutation relations are known and therefore the commutation and anti-commutation relations of the generators are calculated straightforwardly to be:
\begin{equation}
\label{algebra}
\begin{aligned}
&{\left[P_{a}, P_{b}\right]=i \epsilon_{a b c} M_{c}, \quad\left[P_{a}, M_{b}\right]=i \epsilon_{a b c} P_{c}, \quad\left[M_{a}, M_{b}\right]=i \epsilon_{a b c} M_{c},} \\
&\left\{P_{a}, P_{b}\right\}=\frac{1}{2} \delta_{a b} \mathbb{I}, \quad\left\{P_{a}, M_{b}\right\}=\frac{1}{2} \delta_{a b} \gamma_{5}, \quad\left\{M_{a}, M_{b}\right\}=\frac{1}{2} \delta_{a b} \mathbb{I}, \\
&{\left[\gamma_{5}, P_{a}\right]=\left[\gamma_{5}, M_{a}\right]=0, \quad\left\{\gamma_{5}, P_{a}\right\}=2 M_{a}, \quad\left\{\gamma_{5}, M_{a}\right\}=2 P_{a}}.
\end{aligned}
\end{equation}
In turn, having at hand the above relations, one may proceed with the construction of the noncommutative gauge theory by considering the covariant coordinate\footnote{In the particle models setting it was denoted by $\phi_a$ while now $\mathcal{X}_a$ is used.}:
\begin{equation}
    \mathcal{X}_{\mu}=\delta_{\mu}{}^a X_a+\mathcal{A}_{\mu}\,,
\end{equation}
where $\mathcal{A}_{\mu}$ is the gauge connection which expands on the various generators as $\mathcal{A}_{\mu}=\mathcal{A}_{\mu}^{I}(X)\otimes T^{I}$, where $T^{I}$ denotes an arbitrary generator, therefore $I=1,\dots,8$. Also, $\mathcal{A}_{\mu}^I$ are the gauge fields taking values in the algebra of $U(2)\times U(2)$ but also they depend on the matrix-represented coordinates of the underlying space, that is why in the following decomposition of the gauge connection:
\begin{equation}
\mathcal{A}_{\mu}(X)=e_{\mu}{}^{a}(X) \otimes P_{a}+\omega_{\mu}{}^{a}(X) \otimes M_{a}+A_{\mu}(X) \otimes i \mathbb{I}+\tilde{A}_{\mu}(X) \otimes \gamma_{5}
\end{equation}
and therefore of the covariant coordinate:
\begin{equation}
\label{expansX}
    \mathcal{X}_{\mu}=X_{\mu}\otimes i\mathbb{I}+e_{\mu}{}^{a}(X) \otimes P_{a}+\omega_{\mu}{}^{a}(X) \otimes M_{a}+A_{\mu}(X) \otimes i \mathbb{I}+\tilde{A}_{\mu}(X) \otimes \gamma_{5}\,,
\end{equation}
the tensor product is present. In a similar way, the parameter of the gauge parameter, $\varepsilon(X)$, expands on the generators as an element of the gauge algebra:
\begin{equation}
    \varepsilon(X)=\xi^{a}(X) \otimes P_{a}+\lambda^{a}(X) \otimes M_{a}+\varepsilon_{0}(X) \otimes i \mathbb{I}+\tilde{\varepsilon}_{0}(X) \otimes \gamma_{5}\,.
\end{equation}
Having the necessary information at hand, the transformation rules of the various gauge fields are calculated in a straightforward way, taking also into consideration equations \eqref{deltaAlpha} and \eqref{anticom}:
\begin{equation}
\begin{aligned}
    \delta e_{\mu}{}^{a}=
        &-i\left[X_{\mu}+A_{\mu}, \xi^{a}\right]+\frac{i}{2}\left\{\xi_{b}, \omega_{\mu c}\right\} \epsilon^{a b c}+\frac{i}{2}\left\{\lambda_{b}, e_{\mu c}\right\} \epsilon^{a b c}\\
        &+i\left[\varepsilon_{0}, e_{\mu}{}^{a}\right]+\left[\lambda^{a}, \tilde{A}_{\mu}\right]+\left[\tilde{\varepsilon}_{0}, \omega_{\mu}{}^{a}\right],\\
    \delta \omega_{\mu}{}^{a}=
        &-i\left[X_{\mu}+A_{\mu}, \lambda^{a}\right]+\frac{i}{2}\left\{\xi_{b}, e_{\mu c}\right\} \epsilon^{a b c}+\frac{i}{2}\left\{\lambda_{b}, \omega_{\mu c}\right\} \epsilon^{a b c}\\
        &+i\left[\varepsilon_{0}, \omega_{\mu}{}^{a}\right]+\left[\xi^{a}, \tilde{A}_{\mu}\right]+\left[\tilde{\varepsilon}_{0}, e_{\mu}{}^{a}\right],\\
    \delta A_{\mu}=
        &-i\left[X_{\mu}+A_{\mu}, \varepsilon_{0}\right]-\frac{i}{4}\left[\xi^{a}, e_{\mu a}\right]-\frac{i}{4}\left[\lambda^{a}, \omega_{\mu a}\right]-i\left[\tilde{\varepsilon}_{0}, \tilde{A}_{\mu}\right],\\
    \delta \tilde{A}_{\mu}=
        &-i\left[X_{\mu}+A_{\mu}, \tilde{\varepsilon}_{0}\right]+\frac{1}{4}\left[\xi^{a}, \omega_{\mu a}\right]+\frac{1}{4}\left[\lambda^{a}, e_{\mu a}\right]+i\left[\varepsilon_{0}, \tilde{A}_{\mu}\right].
\end{aligned}
\end{equation}
Before we move on to the dynamics of the theory, it is meaningful to make a small detour commenting on the above transformations and what happens when a commutative limit is considered. In this condition, the two Abelian gauge fields introduced for reasons of noncommutativity are taken out of the picture, the conventional derivation is recovered $[X_{\mu},f]\rightarrow -i\partial_{\mu}f$ and the commutators vanish leading to:
\begin{align}
\label{de}
    \delta e_{\mu}{}^a=-\partial_{\mu}\xi^a-\epsilon^{abc}(-i\xi_b\omega_{\mu c}-i\lambda_b e_{\mu c}) \\
\label{domega}
    \delta \omega_{\mu}{}^a=-\partial_{\mu}\lambda^a-\epsilon^{abc}(-i\lambda_b\omega_{\mu c}-i\xi_b e_{\mu c})\,.
\end{align}
The above results are identical to the ones obtained in the conventional gauge-theoretic approach of three-dimensional GR up to some rescalings of generators, gauge fields and parameters in the case that a cosmological constant is present, therefore, switching off noncommutativity leads to recovering the results of the corresponding gauge-theoretic approach in the commutative case.

Getting back in track, in order to propose an action, the corresponding field strength (curvature) tensor of the theory has to be written down. This is achieved by considering the anticipated formula of the commutator of the covariant coordinates but augmented by an extra term, that is linear as the right hand side of the commutation relation of the coordinates \eqref{coordcomm}, the presence of which is necessary in order that it transforms covariantly:
\begin{equation}
\label{defR}
    \mathcal{R}_{\mu \nu}(X)=[\mathcal{X}_{\mu}, \mathcal{X}_{\nu}]-i\lambda\epsilon_{\mu \nu \rho}\mathcal{X}^{\rho}.
\end{equation}
The above tensor expands on the various generators of the algebra as:
\begin{equation}
\label{expansR}
    \mathcal{R}_{\mu \nu}(X)=T_{\mu \nu}{}^{a}(X) \otimes P_{a}+R_{\mu \nu}{}^{a}(X) \otimes M_{a}+F_{\mu \nu}(X) \otimes i \mathbb{I}+\tilde{F}_{\mu \nu}(X) \otimes \gamma_{5}.
\end{equation}
Relating the various relations and definitions, \eqref{expansX}, \eqref{defR} and \eqref{expansR}, the components of the curvature tensor are obtained
and the commutative limit leads to the corresponding relations of the conventional gauge-theoretic approach (see the detailed expressions and discussion in \cite{Chatzistavrakidis:2018vfi}).

\subsection*{\textit{The action for a three-dimensional fuzzy gravity}}
To conclude the study of the three-dimensional case, the action that is proposed is aligned to the ordinary case, in which a Chern-Simons type is employed, that is:
\begin{equation}
S_{0}=\frac{1}{g^{2}} \operatorname{Tr}\left(\frac{i}{3} \epsilon^{\mu \nu \rho} X_{\mu} X_{\nu} X_{\rho}-m^{2} X_{\mu} X^{\mu}\right),\label{rlambdaaction}
\end{equation}
which, following its variation, leads to the field equation:
\begin{equation}
\left[X_{\mu}, X_{\nu}\right]+2 i m^{2} \epsilon_{\mu \nu \rho} X^{\rho}=0\,.
\end{equation}
Variation gives the field equations which admit the background space as a solution for $2m^2=-\lambda$, as expected. Introducing the gauge fields into the above picture by replacing the coordinates with their covariant counterparts, one ends up with:
\begin{equation}
S=\frac{1}{g^{2}} \operatorname{Tr}\operatorname{tr}\left(\frac{i}{3} \epsilon^{\mu \nu \rho} \mathcal{X}_{\mu} \mathcal{X}_{\nu} \mathcal{X}_{\rho}+\frac{\lambda}{2} \mathcal{X}_{\mu} \mathcal{X}^{\mu}\right)\,,
\end{equation}
in which $\operatorname{Tr}$ is effectively the integration operator coordinates and $\operatorname{tr}$ acts on the generators. Taking into consideration the non-vanishing traces $\operatorname{tr}(P_a P_b)=\delta_{a b}, \operatorname{tr}(M_a M_b)=\delta_{a b}$ and, finally, performing the variation with respect to the various gauge fields, the field equations are obtained:
\begin{equation}
    T_{\mu \nu}{}^a=0,~~ R_{\mu \nu}{}^a=0,~~ F_{\mu \nu}=0,~~ \tilde{F}_{\mu \nu}=0\,.
\end{equation}

\section{Fuzzy gravity in four dimensions}
In this section a four-dimensional noncommutative gravitational model as a (noncommutative) gauge theory is constructed. Besides the fact that the four-dimensional case is rather more interesting than the three-dimensional one, the general setting and the methodology that are encountered in the former will be essentially an extension of those of the latter. In this case too, the starting point is the determination of the background space. 

\subsection*{\textit{An approach to the fuzzy Four-Sphere}}
The background noncommutative space that has been chosen to accommodate the construction of the gauge theory is the four-dimensional version of the fuzzy sphere, that is the fuzzy four-sphere, $S_F^4$ \footnote{In this case too, the signature of the space is Euclidean but a construction for the Lorentzian case in which the fuzzy de Sitter space, $dS_F^4$, is employed is possible.}.
In order to find the definition of the space in this four-dimensional case, it would be tempting to attempt a straightforward translation of the fuzzy two-sphere defining relations (commutation and radius) to the four dimensions. Quickly recalling, the coordinates in the fuzzy sphere case came from rescaling the angular momentum operators, i.e. the generators of the isometry group $SO(3)$. If we tried, in the four-dimensional case, to identify four out of the ten generators of the corresponding isometry group $SO(6)$ to the coordinates, we would fall into a dead end because there is no subalgebra that would be nicely closing, or, in other words, the virtue the fuzzy two-sphere had but the four-sphere does not is that of covariance. However, covariance is an essential property of the background fuzzy space, therefore either it should be abandoned or find a way and restore covariance. Selecting the second option, a way to achieve the restoration of the covariance is to consider the coordinates as a subset of a larger group in which the corresponding subalgebra will close \cite{Heckman:2014xha}. The minimal cost one would pay to achieve that is to go to $SO(6)$ \cite{Manolakos:2019fle}, \cite{Manolakos:2021rcl} with its generators obeying the following algebra:
\begin{equation}[J_{AB},J_{CD}]=i(\delta_{AC}J_{BD}+\delta_{BD}J_{AC}-\delta_{BC}J_{AD}-\delta_{AD}J_{BC})\,.
\end{equation}
Adopting an $SO(4)$ notation, after the introduction of a length parameter, $\lambda$, the generators become:
\begin{equation}
    J_{\mu\nu}=\frac{1}{\hbar}\Theta_{\mu\nu}, \qquad J_{\mu 5}=\frac{1}{\lambda}X_5, \qquad J_{\mu 6}=\frac{\lambda}{2\hbar}P_\mu,\qquad J_{56}=\frac{1}{2}h,
\end{equation}
where $\mu,\nu=1,\ldots,4$, $X_\mu, P_\mu$ and $\Theta_{\mu \nu}$ denote the coordinates, momenta and noncommutativity tensor, respectively, and $h$ is an operator bearing information of the radius constraint \cite{Manolakos:2021rcl}. Taking these redefinitions into consideration, the above commutation relation consisting the algebra of $SO(6)$ becomes:
\begin{align}
    [X_\mu ,X_\nu ]=&i\frac{\lambda^2}{\hbar}\Theta_{\mu \nu}, \quad [P_\mu, P_\nu]=4i\frac{\hbar}{\lambda^2}\Theta_{\mu \nu} \quad [P_\mu, h]=4i\frac{\hbar}{\lambda^2}X_{\mu}, \quad [X_\mu, P_\nu]=i\hbar \delta_{\mu \nu} h, \quad [X_\mu, h]=i\frac{\lambda^2}{\hbar} P_\mu\,.
\end{align}
The first commutation relation in the above set of relations corresponds to the defining one of the background fuzzy space, obviously closing into an $SO(4)$ subalgebra of the total $SO(6)$, as aimed. The rest of the commutation relations that come from the decomposition of the initial commutation relation of the $SO(6)$ algebra are the spacetime transformations:
\begin{align}
    [\Theta_{\mu \nu}, \Theta_{\rho \sigma}]&=i\hbar(\delta_{\mu \rho}\Theta_{\nu \sigma}+\delta_{\nu \sigma}\Theta_{\mu \rho}-\delta_{\nu \rho}\Theta_{\mu \sigma}-\delta_{\mu \sigma}\Theta_{\nu \rho})\,,\quad [h, \Theta_{\mu \nu}]=0\,,\\
    [X_\mu ,\Theta_{\nu \rho}]&=i\hbar (\delta_{\mu \rho} X_\nu - \delta_{\mu \nu} X_\rho )\,,\quad 
    [P_\mu ,\Theta_{\nu \rho}]=i\hbar (\delta_{\mu \rho} P_\nu - \delta_{\mu \nu} P_\rho )\,,
\end{align}
where the first one is the $SO(4)$ subalgebra (four-dimensional rotations) and the second one shows how the coordinate vector transforms under these rotations, that is as vectors, validating the covariance property in a rather pronounced way. 
  
\subsection*{\textit{On the Gauge Group }}

As noted throughout the text so far, when constructing a noncommutative gauge theory, the anticommutators become important and, therefore, the initial algebra of the gauge group, here the isometry group $SO(5)$, expands. For the sake of a finite enlargement of the algebra, the representation of the generators gets fixed and, furthermore, for a minimal one, that representation is chosen to be the four-dimensional one and, consequently, the algebra of the gauge group extends to $SO(6)\times U(1)$. The four-dimensional matrices representing the sixteen generators are given in terms of combinations of the the (Euclidean) $4\times 4$ gamma matrices as:
\begin{equation}
M_{ab}=-\frac{i}{4}[\Gamma_a ,\Gamma_b ]\,,\quad K_a=\frac{1}{2}\Gamma_a\,,\quad P_a=-\frac{i}{2}\Gamma_a \Gamma_5\,, \quad D=-\frac{1}{2}\Gamma_5\,, \quad \mathbb{I}_{4}\,,
\end{equation}
which, satisfy the well-known anticommutation relation $\{ \Gamma_a , \Gamma_b \}=2\delta_{ab}\mathbb{I}_{4}$, where $a,b=1,...,4$ and $\Gamma_5=\Gamma_1 \Gamma_2 \Gamma_3 \Gamma_4$.
Therefore, one can now write down a complete list of  commutation and anticommutation relations of all the generators:
\begin{equation}\label{commanticomm}
    \begin{aligned}
        &{\left[K_{a}, K_{b}\right]=i M_{a b}, \quad\left[P_{a}, P_{b}\right]=i M_{a b}},
        {\left[P_{a}, D\right]=i K_{a}, \quad\left[K_{a}, P_{b}\right]=i \delta_{a b} D, \quad\left[K_{a}, D\right]=-i P_{a}}, {\quad\left[D, M_{a b}\right]=0}, \\
        &{\left[K_{a}, M_{b c}\right]=i\left(\delta_{a c} K_{b}-\delta_{a b} K_{c}\right)}, 
        {\quad\left[P_{a}, M_{b c}\right]=i\left(\delta_{a c} P_{b}-\delta_{a b} P_{c}\right)}, {\quad\left\{P_{a}, K_{b}\right\}=\left\{M_{a b}, D\right\}=-\frac{\sqrt{2}}{2} \epsilon_{a b c d} M_{c d}},\\
        &{\left[M_{a b}, M_{c d}\right]=i\left(\delta_{a c} M_{b d}+\delta_{b d} M_{a c}-\delta_{b c} M_{a d}-\delta_{a d} M_{b c}\right)},
        {\quad\left\{M_{a b}, M_{c d}\right\}=\frac{1}{8}\left(\delta_{a c} \delta_{b d}-\delta_{b c} \delta_{a d}\right) \mathbb{I}_{4}-\frac{\sqrt{2}}{4} \epsilon_{a b c d} D},\\ 
        &{\left\{M_{a b}, K_{c}\right\}=\sqrt{2} \epsilon_{a b c d} P_{d}, \quad\left\{M_{a b}, P_{c}\right\}=-\frac{\sqrt{2}}{4} \epsilon_{a b c d} K_{d}}, {\quad\left\{K_{a}, K_{b}\right\} =\frac{1}{2} \delta_{a b} \mathbb{I}_{4}}, {\quad\left\{P_{a}, P_{b}\right\}=\frac{1}{8} \delta_{a b} \mathbb{I}_{4}},\\ &{\left\{K_{a}, D\right\}=\left\{P_{a}, D\right\}=0}.
    \end{aligned}
\end{equation}
Having determined the gauge group, the representation and, therefore, the above relations, it is now possible to move on with the construction of the corresponding gauge theory.  
\subsection*{\textit{Action and Equations of Motion}}
From here on, there are two equivalent ways to proceed, the first one is to follow the straightforward methodology, that is the introduction of the covariant coordinate and the gauge fields, the calculation of the field strength tensor and then the determination of an action. Here, we choose the alternative route, in which an initial topological action is proposed and the gauge fields, covariant coordinate and field strength tensor are involved as a consequence of the introduction of dynamics. According to the above, the starting action is\footnote{Despite the first term of this action, eq.\eqref{topolaction}, $\operatorname{Tr}[X_\mu,X_\nu][X_\rho,X_\sigma]\epsilon^{\mu\nu\rho\sigma}$ vanishes identically, it remains present for later use.}:
\begin{equation}
\label{topolaction}    \mathcal{S}=\operatorname{Tr}\left(\left[X_{\mu}, X_{\nu}\right]-\kappa^{2} \Theta_{\mu \nu}\right)\left(\left[X_{\rho}, X_{\sigma}\right]-\kappa^{2} \Theta_{\rho \sigma}\right) \epsilon^{\mu \nu \rho \sigma}\,,
\end{equation}
with the following field equations:
\begin{equation}
    \epsilon^{\mu \nu \rho \sigma}\left[X_{\nu},\left[X_{\rho}, X_{\sigma}\right]-\kappa^{2} \Theta_{\rho \sigma}\right]=0\,,\,\, \epsilon^{\mu \nu \rho \sigma}\left(\left[X_{\rho}, X_{\sigma}\right]-\kappa^{2} \Theta_{\rho \sigma}\right)=0\,,
\end{equation}
which are obtained after varying with respect to $X$ and $\Theta$. From the second equation, in case $\kappa^2=\frac{i\lambda^2}{\hbar}$, the defining relation of noncommutativity of the space is recovered and, therefore, the first equation is automatically satisfied. Dynamics to the above action are introduced after the consideration of gauge fields as fluctuations of $X$ and $\Theta$:
\begin{equation}
\begin{gathered}
    \mathcal{S}=\operatorname{Trtr} \epsilon^{\mu \nu \rho \sigma}\left(\left[X_{\mu}+A_{\mu}, X_{\nu}+A_{\nu}\right]-\kappa^{2}\left(\Theta_{\mu \nu}+\mathcal{B}_{\mu \nu}\right)\right)\\
   \qquad \qquad \qquad \qquad \qquad \cdot \left(\left[X_{\rho}+A_{\rho}, X_{\sigma}+A_{\sigma}\right]-\kappa^{2}\left(\Theta_{\rho \sigma}+\mathcal{B}_{\rho \sigma}\right)\right)\,,
\end{gathered}
\end{equation}
where $\operatorname{tr}$ denotes the trace over the generators of the algebra of the gauge group. The above expression of the action can resemble formally the (ordinary) four-dimensional Chern-Simons action, after the following identifications:   
\begin{itemize}
    \item The covariant coordinate: $\mathcal{X}_{\mu}\equiv X_{\mu}+A_{\mu}$, where $A_{\mu}$ is identified as the gauge connection:
    \[A_\mu = e_\mu^{~a}\otimes P_a + \omega_\mu^{~ab}\otimes M_{ab} + b_\mu^{~a}\otimes K_a + \tilde{a}_\mu\otimes D + a_\mu \otimes \mathbb{I}_{4}\]
    \item The covariant noncommutative tensor $\hat{\Theta}_{{\mu}{\nu}}\equiv\Theta_{{\mu}{\nu}}+\mathcal{B}_{{\mu}{\nu}}$, where $\mathcal{B}_{{\mu}{\nu}}$ is a 2-form field
    \item The (covariant) field strength tensor of the theory: $\mathcal{R}_{\mu \nu}\equiv\left[\mathcal{X}_{\mu}, \mathcal{X}_{\nu}\right]-\kappa^2 \hat{\Theta}_{\mu \nu}$, which, as an element of the gauge algebra, can be decomposed in various component tensors\footnote{The detailed expressions of the components of the field strength tensors as well as the transformations of the gauge fields can be found in the original papers \cite{Manolakos:2019fle, Manolakos:2021rcl}.}:
\begin{equation}
    \mathcal{R}_{\mu \nu}(X)=\tilde{R}_{\mu \nu}{}^{a} \otimes P_{a}+R_{\mu \nu}{}^{a b} \otimes M_{a b}+R_{\mu \nu}{}^{a} \otimes K_{a}+\tilde{R}_{\mu \nu} \otimes D+R_{\mu \nu} \otimes \mathbb{I}_{4}\,.\label{fieldstrengthdecomp}
\end{equation}
\end{itemize}
In turn, using $\kappa^2=\frac{i\lambda^2}{\hbar}$, the expression of the action becomes:
\begin{equation}
\label{dynamicaction}
\mathcal{S}=\operatorname{Trtr}\left(\left[\mathcal{X}_{\mu}, \mathcal{X}_{\nu}\right]-\frac{i \lambda^{2}}{\hbar} \hat{\Theta}_{\mu \nu}\right)\left(\left[\mathcal{X}_{\rho}, \mathcal{X}_{\sigma}\right]-\frac{i \lambda^{2}}{\hbar} \hat{\Theta}_{\rho \sigma}\right) \epsilon^{\mu \nu \rho \sigma}:=\operatorname{Trtr} \mathcal{R}_{\mu \nu} \mathcal{R}_{\rho \sigma} \epsilon^{\mu \nu \rho \sigma}
\end{equation}
and variations lead once again to two kinds of field equations:
\begin{equation}
    \epsilon^{\mu \nu \rho \sigma} \mathcal{R}_{\rho \sigma}=0\,,\,\, \epsilon^{\mu \nu \rho \sigma}\left[\mathcal{X}_{\nu}, \mathcal{R}_{\rho \sigma}\right]=0\,,
\end{equation}
which are understood as the vanishing of the field strength tensor and a noncommutative analogue of the Bianchi identity, respectively. 
\subsection*{\textit{Symmetry Breaking of the Action}}
The above action of Chern-Simons type  \eqref{dynamicaction}, is invariant under $SO(6)\times U(1)$ gauge symmetry which is a rather big amount of symmetry as a result of the enlargement of the group. For reasons of compatibility of the two frameworks (commutative-noncommutative), especially when the commutative limit is considered, the above gauge group can break down to a smaller one either by imposing certain constraints \cite{Manolakos:2019fle} or by introducing an auxiliary gauge field and letting it acquire a vev \cite{Manolakos:2021rcl}. According to the second recipe, the action \eqref{dynamicaction} becomes:  
\begin{equation}
\label{modaction}
    \mathcal{S}=\operatorname{Trtr}_{G} \lambda \Phi(X) \mathcal{R}_{\mu \nu} \mathcal{R}_{\rho \sigma} \epsilon^{\mu \nu \rho \sigma}+\eta\left(\Phi(X)^{2}-\lambda^{-2} \mathbf{I}_{N} \otimes \mathbf{I}_{4}\right)\,,
\end{equation}
where $\eta$ is a Lagrange multiplier and $\lambda$ is a parameter of dimension of length. On-shell it holds that:
\begin{equation}
\Phi^2(X)=\lambda^{-2}\mathbb{I}_{N}\otimes\mathbb{I}_{4}\,, 
\end{equation}
which means that the initial action remains effectively untouched. Now, let the scalar field, $\Phi$, consist exclusively of the symmetric component, therefore it is expressed as:
\[
\Phi(X)=\tilde{\phi}^{a}(X) \otimes P_{a}+\phi^{a}(X) \otimes K_{a}+\phi(X) \otimes \mathbb{I}_{4}+\tilde{\phi}(X) \otimes D\,.
\] 
The symmetry breaking takes place when the scalar field is gauge-fixed along the direction of the $D$ generator, that is:
\[
    \Phi(X)=\tilde{\phi}(X) \otimes D|_{\tilde{\phi}=-2 \lambda^{-1}}=-2 \lambda^{-1} \mathbf{I}_{N} \otimes D.
\]
Putting the above value of the scalar field into the action \eqref{modaction} and perform the trace over the generators, the action becomes:
\begin{equation}
\label{brokenaction}
\mathcal{S}_{\text{br}}=\operatorname{Tr}\left(\frac{\sqrt{2}}{4} \epsilon_{a b c d} R_{\mu \nu}{}^{a b} R_{\rho \sigma}{}^{c d}-4 R_{\mu \nu} \tilde{R}_{\rho \sigma}\right) \epsilon^{\mu \nu \rho \sigma}\,.
\end{equation}
The consideration that the scalar field consists only of the symmetric part of the algebra of $SO(6)$ and implying at the same time that it is not charged under the $U(1)$, the gauge symmetry of the resulting action is the $SO(4)\times U(1)$. Specifically, the generators that remain unbroken in the above mechanism are: a) $P_a$ which leads to the torsionless condition which, in turn results to relation between the gauge fields $\omega, e$ and $\tilde{a}$, b) $K_a$ which lead to the condition $R_{\mu \nu}{}^{a}=0$, which in turn implies a relation of proportionality between $e, b$ gauge fields and c) $D$ which requires $\tilde a_\mu=0$ \cite{2003}. Therefore, the remnant symmetry of the spontaneously broken theory is $SO(4)\times U(1)$ and the independent fields are $e$ and $a$. Last, the resulting expression of the component tensor $R_{\mu \nu}{}^{ab}$, after taking into consideration the following results of the conditions from the breaking $\tilde{a}_\mu=0$ and $b_\mu{}^a=\frac{i}{2}e_\mu{}^a$, is:
\begin{align}
R_{\mu \nu}{}^{a b}=&\left[X_{\mu}+a_{\mu}, \omega_{\nu}{}^{a b}\right]-\left[X_{\nu}+a_{\nu}, \omega_{\mu}{}^{a b}\right]+i\left\{\omega_{\mu}{}^{a c}, \omega_{\nu~c}^{~b}\right\}-i\left\{\omega_{\mu}{}^{b c}, \omega_{\nu c}{}^{a}\right\} \\
&+\frac{3 i}{8}\left\{e_{\mu}{}^{a}, e_{\nu}{}^{b}\right\}-\frac{i \lambda^{2}}{\hbar} B_{\mu \nu}{}^{a b}.
\end{align}
\section{A quick glimpse of the metric}
Working with the (noncommutative) gauge-theoretic approach to gravity, the notion of the metric is not explicitly present. Instead, the quantity that relates the above with the geometric approach is the vierbein which has been introduced in the construction as a gauge field related to the generators of the translations. In the conventional framework, the above two are related through the well-known "square root of the metric" equation which, according to ref \cite{2003}, in the noncommutative regime and specifically in the star-product realization, it reads:
$$g_{\mu\nu}=e_\mu^{~a}\star e_{\nu a}\,,$$
where, in the same source, it is argued that the above relation leads to theories of complex gravity. In order to translate the above to the matrix realization of noncommutativity which is of our interest, we consider the following version of the above relation:
\begin{equation}
    g_{\mu\nu}=(e_\mu^{~a}\otimes P_a) (e_\nu^{~b}\otimes P_b) = \frac{1}{2}e_\mu^{~a}e_\nu^{~b}\otimes iM_{ab}+\frac{1}{2}e_\mu^{~a}e_\nu^{~b}\otimes \frac{1}{8}\delta_{ab}\mathbb{I}_4\label{metric1}\equiv G_{\mu\nu}+iB_{\mu\nu}\,,
\end{equation}
where the commutation and anticommutation relations of the $P$ generators as found in eq.\eqref{commanticomm} have been used. By inspection, both real and imaginary parts are encountered, therefore it can be deduced that the metric is complex in this case as well. Also, considering the $g_{\mu\nu}$ to be hermitian
leads to the understanding that $G_{\mu\nu}$ is symmetric while the $B_{\mu\nu}$ is antisymmetric under the exchange of the spacetime indices $\mu,\nu$. Therefore, comparing to the above expression of the metric, \eqref{metric1}, the following identification is achieved:
\begin{align}
    G_{\mu\nu}&=\frac{1}{16}e_\mu^{~a}e_{\nu a}\otimes \mathbb{I}_4=\frac{1}{32}\{e_{\mu}^{~a},e_{\nu a}\}\otimes \mathbb{I}_4\,,\label{metric2}\\
    B_{\mu\nu}&=\frac{1}{2}e_\mu^{~a}e_\nu^{~b}\otimes M_{ab}=\frac{1}{4}[e_{\mu}^{~a},e_\nu^{~b}]\otimes M_{ab}\,.
\end{align}
It has been routed for future projects to delve more into the above observations regarding the metric. 

\section{Noncommutative gravity on fuzzy 3-sphere}
Although the three-dimensional case was examined earlier, here is the right place to examine the case of a noncommutative gravitational model in three-dimensions in case the background space in not $\mathbb{R}_\lambda^3$ like in the previous three-dimensional case but the fuzzy 3-sphere, $S_{F}^{3}$. It is examined now because the methodology of building this version of the fuzzy 3-sphere is a lower-dimensional analogue of the construction of the fuzzy four-sphere as it got realized in the previous section. Like the four-dimensional fuzzy sphere, this space is also both of Lie and Moyal type since the noncommutative tensor is both constant and at the same time a generator of the isometry group. So, following the same methodology as in \cite{Manolakos:2021rcl}, starting from the isometry group $SO(4)$ an extension is required for reasons of covariance, therefore the $SO(5)$ is eventually used. Then, a 2-step procedure is followed in which, in the first step, the space is manifested as an embedding in the four-dimensional Euclidean spacetime, while in the second step in the three-dimensional one. The resulting space will be a realization of the fuzzy 3-sphere and will finally be employed for the construction of a gravity model on it. 

\subsection*{First step: $SO(5) \supset SO(4)$}
We start with the $SO(5)$ group which has $10$ hermitian generators, $J_{MN}$, where $M, N=0,\ldots, 4$ which obey:
\begin{equation}
\left[J_{M N}, J_{P \Sigma}\right]=i\left(\delta_{M P} J_{N \Sigma}+\delta_{N \Sigma} J_{M P}-\delta_{N P} J_{M \Sigma}-\delta_{M \Sigma} J_{N P}\right),
\end{equation}
where $\delta_{MN}$ is the 5-dim Kronecker delta. The $SO(5) \supset SO(4)$ decomposition yields:
\begin{itemize}

\item For $M=m, N=m, P=r, \Sigma=s$:\quad
$\left[J_{m n}, J_{r s}\right]=i\left(\delta_{m r} J_{n s}+\delta_{n s} J_{m r}-\delta_{n r} J_{m s}-\delta_{m s} J_{n r}\right)\,.$

\item For $M=m, N=4, P=r, \Sigma=4$:\quad
$\left[J_{m 4}, J_{n 4}\right]=i J_{m n}\,.$
\item For $M=m, N=n, P=r, \Sigma=4$:\quad $\left[J_{m n}, J_{r 4}\right]=i\left(\delta_{m r} J_{n 4}-\delta_{n r} J_{m 4}\right)\,,$
\end{itemize}
where $m, n, r, s=0, \ldots, 3$. Setting $\Theta_{m n} \equiv \hbar J_{m n}, X_{m} \equiv \lambda J_{m 4}$, where $\lambda$ is a parameter of dimension of length, the above
commutation relations become:
\begin{equation}\label{from}
\begin{aligned}
{\left[\Theta_{m n}, \Theta_{r s}\right] } &=i \hbar\left(\delta_{m r} \Theta_{n s}+\delta_{n s} \Theta_{m r}-\delta_{n r} \Theta_{m s}-\delta_{m s} \Theta_{n r}\right), \\
{\left[\Theta_{m n}, X_{r}\right] } &=i \hbar\left(\delta_{m r} X_{n}-\delta_{n r} X_{m}\right), \quad
{\left[X_{m}, X_{n}\right] } =i \frac{\lambda^{2}}{\hbar} \Theta_{m n}\,.
\end{aligned}
\end{equation}
Also, along the lines of the fuzzy 2-sphere, the quadratic Casimir element is related to the radius constraint:
\begin{equation}\label{cas}
\begin{aligned}
&C_{2}^{S O(5)}=-\frac{1}{2} \operatorname{Tr} J^{2}=\frac{1}{2} J_{M N} J^{M N}=\frac{1}{2} J_{m n} J^{m n}+J_{m 4} J^{m 4} \Rightarrow\\
&X^{m} X_{m}=\lambda^{2}\left(C_{2}^{S O(5)}-C_{2}^{S O(4)}\right) \equiv r^{2} ,
\end{aligned}
\end{equation}
where $r^2\equiv \lambda^2L$, and where it has been assumed that the $SO(5)$ and $SO(4)$ generators are of the same spin representation, $L$.
\subsection*{Second step: $SO(4) \supset SO(3)$}
From the first equation of \eqref{from}, if $\mu, \nu, \rho, \sigma=0, \ldots, 2$:
\begin{itemize}
\item  For $m=\mu, n=\nu, r=\rho, s=\sigma$, where $\mu, \nu, \rho, \sigma=0, \ldots, 2$ :
$$
\left[\Theta_{\mu \nu}, \Theta_{\rho \sigma}\right]=i \hbar\left(\delta_{\mu \rho} \Theta_{\nu \sigma}+\delta_{\nu \sigma} \Theta_{\mu \rho}-\delta_{\nu \rho} \Theta_{\mu \sigma}-\delta_{\mu \sigma} \Theta_{\nu \rho}\right)\,.
$$
\item  For $m=\mu, n=\nu, r=\rho, s=3$: \quad
$
\left[\Theta_{\mu \nu}, \Theta_{\rho 3}\right]=i \hbar\left(\delta_{\mu \rho} \Theta_{\nu 3}-\delta_{\nu \rho} \Theta_{\mu 3}\right)\,.
$
\item  For $m=\mu, n=3, r=\rho, s=3$:\quad
$
\left[\Theta_{\mu 3}, \Theta_{\rho 3}\right]=i \hbar \Theta_{\mu \rho}\,.
$
\end{itemize}
From the second equation of \eqref{from}:
\begin{itemize}
\item For $m=\mu, n=\nu, r=\rho$:\quad
$
\left[\Theta_{\mu \nu}, X_{\rho}\right]=i \hbar\left(\delta_{\mu \rho} X_{\nu}-\delta_{\nu \rho} X_{\mu}\right)\,.
$
\item For $m=\mu, n=\nu, r=3$:\quad
$\left[\Theta_{\mu \nu}, X_{3}\right]=0\,.$
\item For $m=\mu, n=3, r=3$:\quad
$
\left[\Theta_{\mu 3}, X_{3}\right]=-i \hbar X_{\mu}\,.
$
\item For $m=\mu, n=3, r=\rho$ :
$
\left[\Theta_{\mu 3}, X_{\rho}\right]=i \hbar \delta_{\mu \rho} X_{3}\,.
$
\end{itemize}
From the third equation of \eqref{from}:
\begin{itemize}
\item For $m=\mu, n=\nu$:\quad
$
\left[X_{\mu}, X_{\nu}\right] =i \frac{\lambda^{2}}{\hbar} \Theta_{\mu \nu}\,.
$
\item For $m=\mu, n=3$:\quad
$
\left[X_{\mu}, X_{3}\right]=i \frac{\lambda^{2}}{\hbar} \Theta_{\mu 3}\,.
$
\end{itemize}
Setting $P_{\mu} \equiv \frac{1}{\lambda} \Theta_{\mu 3}, h \equiv \frac{1}{\lambda} X_{3}$, the commutation relations regarding all the operators $\Theta_{\mu \nu}, X_{\mu}, P_{\mu}, h$ are:
\begin{equation}\label{al}
\begin{aligned}
&{\left[\Theta_{\mu \nu}, \Theta_{\rho \sigma}\right] } =i \hbar\left(\delta_{\mu \rho} \Theta_{\nu \sigma}+\delta_{\nu \sigma} \Theta_{\mu \rho}-\delta_{\nu \rho} \Theta_{\mu \sigma}-\delta_{\mu \sigma} \Theta_{\nu \rho}\right),
\quad{\left[P_{\mu}, P_{\nu}\right] }=i \frac{\hbar}{\lambda^{2}} \Theta_{\mu \nu}, \quad\left[X_{\mu}, X_{\nu}\right]=i \frac{\lambda^{2}}{\hbar} \Theta_{\mu \nu}, \\
&{\left[P_{\mu}, h\right] } =-i \frac{\hbar}{\lambda^{2}} X_{\mu}, \quad\left[X_{\mu}, h\right]=i \frac{\lambda^{2}}{\hbar} P_{\mu},\quad
{\left[\Theta_{\mu \nu}, P_{\rho}\right] } =i \hbar\left(\delta_{\mu \rho} P_{\nu}-\delta_{\nu \rho} P_{\mu}\right),\\ &\left[\Theta_{\mu \nu}, X_{\rho}\right]=i \hbar\left(\delta_{\mu \rho} X_{\nu}-\delta_{\nu \rho} X_{\mu}\right), \quad{\left[P_{\mu}, X_{\nu}\right] } =i \hbar \delta_{\mu \nu} h, \quad\left[\Theta_{\mu \nu}, h\right]=0\,.
\end{aligned}
\end{equation}
The embedding relation,\eqref{cas}, in $SO(3)$ notation, becomes: 
\begin{equation}
X^{m} X_{m}=r^{2} \Rightarrow X^{\mu} X_{\mu}+X_{3} X^{3}=r^{2} \Rightarrow h=\pm \sqrt{\frac{1}{\lambda^{2}}\left(X_{\mu} X^{\mu}-r^{2}\right)}\,,
\end{equation}
where just like in the case of the fuzzy 4-sphere, $h$ is an operator related to the radius constraint. Also, it is worth noting that in the limit where $h\rightarrow 0$, a new realization of the fuzzy 2-sphere, $S_{F}^{2}$ emerges.
\subsection*{Determination of the action}
This noncommutative space, as a fuzzy version of the 3-sphere, has its isometries parametrized by the $SO(4)$ group, therefore the construction of the corresponding noncommutative gauge theory will coincide to the one presented above in the $\mathbb{R}_\lambda^3$ case. So, according to the results of section 4, the minimum expansion of this group, in order that the resulting operators of the anticommutators to be contained, is again $U(2)\times U(2)$ which consists of the same set of generators satisfying the same algebra as it is written in \eqref{algebra}.
The difference between the two cases of $S_F^3$ and $\mathbb{R}_\lambda^3$ emerges when considering the action. Although, in this case too, an action of three-dimensional Chern-Simons type is used, compared to \eqref{rlambdaaction}, the second term that is related to the kind of noncommutativity is different:
\begin{equation}
\label{topolaction1}
\mathcal{S}=\operatorname{Tr}\left[\epsilon^{\mu \nu \rho}\left(\frac{i}{3}X_\mu X_\nu X_\rho + \kappa^2 \{X_\mu , \Theta_{\nu \rho}\}\right)\right]
\end{equation}
Variation of the above action will lead to the corresponding field equations. It is expected
that our background space should satisfy the derived equations. We consider that, in
principle, $X$ and $\Theta$ are independent fields. Variation with respect to $\Theta$ and $X$ respectively gives:
\begin{equation}\label{1}
\epsilon^{\mu \nu \rho} X_\mu=0\,, \quad \epsilon^{\mu \nu \rho}\left(\left[X_\mu , X_\nu \right]-2i\kappa^2\Theta_{\mu \nu}\right)=0 \,.
\end{equation}
Despite the second field equation is satisfied by the $S_F^3$ when $\kappa^{2}=\frac{ \lambda^{2}}{4\hbar}$, meaning that the background space is indeed a solution, the first one, $X_\mu=0$, is effectively a trivialization. If we move on to examine the case in which gauge fields are introduced as fluctuations, setting $\kappa^{2}=\frac{ \lambda^{2}}{4\hbar}$, introducing a trace over the gauge algebra and making use of similar definitions as the ones above eq.\eqref{fieldstrengthdecomp}, the action given in eq. \eqref{topolaction1} will become:
\begin{equation}
S=\operatorname{Trtr}\left[ \epsilon^{\mu \nu \rho \sigma}\left(\frac{i}{6}\mathcal{X}_\mu \mathcal{R}_{\nu \rho} + \frac{\lambda^2}{4\hbar}\{\mathcal{X}_\mu , \hat{\Theta}_{\nu \rho}\}\right)\right]
\end{equation}
and, in turn, will give the following field equations: 
\begin{equation}\label{fe1}
\epsilon^{\mu \nu \rho}\mathcal{X}_\mu=0\,,\quad \epsilon^{\mu \nu \rho}\mathcal{R}_{\mu \nu}=0\,.
\end{equation}
Again, the second equation is the vanishing of the field strength tensor, $\mathcal{R}_{\mu \nu}=\left[\mathcal{X}_{\mu}, \mathcal{X}_{\nu}\right]- \frac{i\lambda^{2}}{2\hbar} \hat{\Theta}_{\mu \nu}$, which would be an acceptable field equation, but the first one trivializes the model.

Summing up, the above construction of a three-dimensional noncommutative gravity on a fuzzy 3-sphere has a lot in common with the one constructed in section 3 on the $\mathbb{R}_\lambda^3$ space. Nevertheless, there are some differences in the two gauge theories that become manifest when the kind of space manifestly enters in the calculations. The action is again of three-dimensional Chern-Simons type, but involves the $\Theta$ operator instead of only the $X$, \eqref{topolaction1}. This consideration led to the field equations of eq.\eqref{fe1}, which, due to the first one, imply that no interesting physical conclusions are achieved by this model. Maybe the employment of an alternative action would produce more interesting results, leaving this task for a future study.
\section{Unification of conventional and fuzzy four-dimensional gravity with gauge interactions}
As it has been already discussed in the introduction there exists a gauge theoretic construction of GR \cite{Utiyama:1956sy, Kibble:1961ba, Stelle:1979aj, MacDowell:1977jt, Ivanov:1980tw, Ivanov:1981wm, Kibble:1985sn, Kaku:1977pa, Fradkin:1985am, vanproeyen, cham-thesis, Chamseddine:1976bf, Witten:1988hc} in addition to the geometric one, namely the description of GR as a gauge theory of the Lorentz group with the spin connection as the corresponding gauge field which would enter in the action through the corresponding field strength. Usually the dimension of the tangent space is taken to be equal to the dimension of the curved manifold. However the dimension of the tangent group is not necessarily the same as the dimension of the manifold \cite{Weinberg:1984ke}.

In ref \cite{Chamseddine:2016pkx} (see also \cite{Nesti:2009kk}) the authors have considered higher-dimensional tangent spaces in four dimensional space-time and managed in this way to achieve unification of gauge interactions with gravity. The geometric unification of gravity and gauge interactions is realized by writing the action of the full theory in terms only of the curvature invariants of the tangent group, which contain the Yang-Mills actions corresponding to the gauge groups describing in this way together the GR and the internal GUT in a unified manner. The best model found so far that unifies gravity and a chiral GUT is based on $SO(1,13)$ in a 14-dimensional tangent space.

In order to make clear the specific model let us consider the decomposition of $SO(14)$ under the maximal subgroup:
$$ SO(14) \supset SU(2) \times SU(2) \times SO(10)\,,$$
where the compact isomorphic image of $SO(1,3)$, $SO(4) = SU(2) \times SU(2)$ is used for convenience. The decomposition of the adjoint representation under the above subgroups is given by:
$$ 91 = (3,1,1) = (1,3,1) + (1,1,45) + (2,2,10)$$
and of the spinor by:
$$64 = (2,1,16) + (1,2, \bar{16}).$$
Noting that the $(2,1), (1,2)$ under $SU(2) \times SU(2)$ are the L, R handed fermions respectively, one finds that the spinor of $SO(1,13)$ describes two chiral $16_L$ fermionic families (since the $\bar{16}_R = 16_L$). Without going further in the analysis the result is that it is indeed possible to achieve unification of GR with internal symmetries expressed as an $SO(10)$ GUT in this case and the number of families will be even, if more $64$ are added. In ref \cite{Chamseddine:2016pkx} in their analysis preferred to make heavy one of the families by suitable spontaneous symmetry breaking.

Let us first note that the above analysis can be extended using the $SO(18)$ and then consider the decompositions under the following maximal subgroups:
$$SO(18) \supset SO(8) \times SO(10)$$
and then
$$ SO(8) \supset SU(2) \times SP(4) \qquad \text{and} \qquad SP(4) \supset SU(2) \times SU(2)\,.$$
In turn choosing one $SU(2)$ from the $SO(8)$ decomposition and another from the $Sp(4)$ decomposition to form the gauging of the Lorentz group that produces the GR, one can obtain again $SO(10)$ unification with $4$ families from the spinor, $256$ of $SO(18)$.

A similar procedure can be applied starting with $SO(22)$ and the decompositions under the following maximal subgroups: 
$$SO(22) \supset SO(12) \times SO(10)$$
with the decomposition of the adjoint:
$$ 231 = (1,45) + (66,1) + (12,10)$$
and then:
$$SO(12) \supset SU(2) \times SU(2) \times SU(2)\,,$$
with decomposition of the adjoint:
$$66 = (3,1,1) + (1,3,1) + (1,1,3) + (3,3,3) + (5,3,1) + (5,1,3)$$
and of the spinor:
$$1024 = (32,16) + (32',\bar{16})\,.$$
Then choosing the last two $SU(2)$s of the above decomposition to be gauged and produce GR, under which the decomposition of the spinors is:
$$32 = (1,4,1) + (3,2,3) + (5,2,1)$$
$$ 32'= (1,1,4) + (3,3,2) + (5,1,2)$$
one can find that from the spinor of of $SO(22$), 1024 chiral $16_L$ are obtained but with a multiplicity 10, which is far too high to become realistic and can be excluded on this basis.

With the above procedure considered as a possible way to unify gravity with gauge internal interactions, we naturally tried to extend it in our four-dimensional construction of fuzzy gravity in which we had to enlarge the tangent space to $SO(6)$ with the $SO(4)$ taken as the maximal subgroup of $SO(5)$, which in turn was the maximal subgroup of $SO(6)$, i.e. the $SO(4)$ was taken from the following chain of maximal subgroups:
\begin{equation}
\label{label}
SO(6) \supset SO(5) \supset SO(4)\,. 
\end{equation}
In turn the gauging procedure to construct fuzzy gravity led us to consider the $SO(6) \times U(1)$ as the appropriate gauge group. A possible way to proceed in the unification of fuzzy gravity with internal gauge symmetries, as in the previous examples, would be to consider the $SO(16)$ as the unifying group. Then the taking the the maximal subgroups
$$SO(16) \supset SO(6) \times SO(10)$$
under which the adjoint is decomposed as:
$$120 = (15,1) + (1,45) + (6,10)$$
and the spinor as:
$$128 = (4,16) + (\bar{4},\bar{16})$$
we could consider in the gauging of $SO(6) \times U(1)$ to identify e.g. the $U(1)$ with the one resulting from the maximal decomposition of
 $$ SO(10) \supset SU(5) \times U(1)$$
with decomposition of the adjoint as:
$$45 = 1(0) + 10(4) + \bar{10} (-4) + 24(0)\,.$$
The problem arises from the fact that according to the decomposition chain \eqref{label} we have that the spinor (antispinor) of $SO(6)$ decomposes as follows:
$$SO(6) \supset SO(5) \supset SO(4)= SU(2) \times SU(2)$$
$$4  =  4  = (2,1) + (1,2)\,,$$
which means that from the decomposition of the spinor $128$ of $SO(16)$ we obtain:
$$SO(16) \supset SO(6) \times SO(10)\,,$$
$$128 = (4,16) + (\bar{4},\bar{16})\,,$$
which in turn under $SO(4) = SU(2) \times SU(2)$ becomes:
$$SO(16) \supset SU(2) \times SU(2) \times SO(10)$$
$$ 128 = ([(2,1) + (1,2)], 16) +  ([(2,1) + (1,2)], \bar{16})\,,$$
which unfortunately is not a chiral theory. It seems that this is a general feature of our construction, which originates from the chosen chain (3) in our construction. Therefore, this very interesting way of unifying four-dimensional gravity and internal interactions of particle physics does not work in the case of the constructed four-dimensional fuzzy gravity.

\section{Conclusions}
In this review, we presented a panoramic view of the studies on the physical perspective of ours on noncommutative geometry, which have been carried out by various collaborators over the years, updated by meaningful comments and extensions. First, we presented the general approach on how gauge theories are formulated in the noncommutative setting, then, a particle physics model in which noncommutativity becomes manifest through fuzzy extra dimensions and, last, gravitational models in three and four dimensions constructed on background fuzzy spaces as gauge theories. At the end, an attempt of unification between the gravity and GUTs was made, which although successful in the commutative case, it failed to result to a fruitful outcome in the noncommutative (fuzzy) one. Overall, intertwining noncommutativity with physics consists a very promising initiative leading to interesting features and models, as in the quantum gravity realm, as in particle physics and extra dimensions.

\section*{Acknowledgements}
We would like to thank all our collaborators in the projects that are covered in this review, Paolo Aschieri, Danijel Jurman, Athanasios Chatzistavrakidis, Lara Jonke, John Madore and Harold Steinacker. We appreciate the useful discussions with Patricia Vitale, Ali Chamseddine, Denjoe O'Connor, Dimitra Karabali, V. Parameswaran Nair, Maja Buri\'c, Dumitru Ghilencea, Ichiro Oda, Emmanuel Saridakis, and Viatcheslav Mukhanov. Finally, we are grateful to Dieter L\"ust and Harald Grosse for their constant encouragement. 

Affiliated authors of NTUA have been supported by the Basic Research Programme, PEVE2020 of National Technical University of Athens, Greece. One of us (GZ) would like to thank the DFG Exzellenzcluster 2181:STRUCTURES of Heidelberg University, MPP-Munich, A.v.Humboldt Foundation and CERN-TH for support. This work (GM) has been supported by the Polish National Science Center Grant No. 2017/27/B/ST2/02531.

\section*{Data Availability Statement}
Data Availability Statement: No Data associated in the manuscript

\end{document}